\documentclass[12pt]{article}
\usepackage{amsfonts}
\usepackage{psfig}

\newcommand{\beq}{\begin{equation}}
\newcommand{\eeq}{\end{equation}}
\newcommand{\beqn}{\begin{eqnarray}}
\newcommand{\eeqn}{\end{eqnarray}}
\newcommand{\bea}[1]{\beq\begin{array}{#1}}
\newcommand{\eea}{\end{array}\eeq}

\newcommand{\dual}[1]{{}^{*}{#1}}

\newcommand{\diff}{\partial}
\newcommand{\bh}{{\bf H}}
\newcommand{\br}{{\bf r}}
\newcommand{\cC}{{\cal C}}

\newcommand{\NP}[3]{{\it Nucl. Phys. }{\bf #1} (#2) #3}
\newcommand{\NPPS}[3]{{\it Nucl. Phys. Proc. Suppl. }{\bf #1} (#2) #3}
\newcommand{\PL}[3]{{\it Phys. Lett. }{\bf #1} (#2) #3}

\newcommand{\PRep}[3]{{\it Phys. Rep. }{\bf #1} (#2) #3}

\newcommand{\PTPS}[3]{{\it Prog. Theor. Phys. Suppl. }{\bf #1} (#2) #3}

\begin{document}\date{}
\title{Magnetic monopoles, alive.
\vskip-40mm
\rightline{\small ITEP-TH-36/00}
\vskip 40mm
}
\author{
M.N. Chernodub$^{\rm a}$,
F.V.~Gubarev$^{\rm a,b}$,
M.I. Polikarpov$^{\rm a}$, 
V.I.~Zakharov$^{\rm b}$ \\
\\
$^{\rm a}$ {\small\it Institute of Theoretical and  Experimental Physics,}\\
{\small\it B.Cheremushkinskaya 25, Moscow, 117259, Russia}\\
$^{\rm b}$ {\small\it Max-Planck Institut f\"ur Physik,}\\
{\small\it F\"ohringer Ring 6, 80805 M\"unchen, Germany}
}

\maketitle
\thispagestyle{empty}
\setcounter{page}{0}
\begin{abstract}\noindent 
We review recent developments in understanding the physics
of the magnetic mo\-no\-po\-les in unbroken non-Abelian gauge theories.
Since numerical data on the monopoles are accumulated in lattice simulations,
the continuum theory is understood as the limiting case of the lattice formulation.
In this review, written for a volume dedicated to 
the memory of Academician A.B. Migdal,
we emphasize physical effects related to the monopoles.
In particular, we discuss the monopole-antimonopole potential at short and
larger distances as well as a dual formulation of the gluodynamics,
relevant to the physics of the confinement.
\end{abstract}
\newpage

\section{Generalities}

\subsection{Introduction} 

Magnetic monopoles is undoubtedly a fascinating subject.
Not a new one, though. The Dirac magnetic monopole is 70  years old soon \cite{dirac}.
And the first 50 years of development of the theory of the magnetic monopoles were summarized
in an illuminating review by Coleman \cite{coleman}.
Thus, the question may arise
why it is instructive to come back to the monopoles now.

The main development since the Coleman's review is that
monopoles were copiously observed 
(for review and further references see, e.g., \cite{review}) 
and the theory can be now confronted with
the data. True, the monopoles observed are not exactly those
introduced by Dirac, but rather their close akins, that is
monopoles of non-Abelian gauge theories 
(moreover, for the sake of definiteness we concentrate 
on the simplest gauge group, that is $SU(2)$).
Also true, the data are
numerical mostly and obtained on the lattice so that
their interpretation in terms of
the continuum theory may not be so straightforward.
Nevertheless, it is a direct challenge to theory to explain the 
ample data on the magnetic monopoles which have already been accumulated
in the lattice simulations.

Moreover, the issue of the so to say lattice monopoles is very much 
rich and varied  by itself. Let us mention here three topics:

(a) The numerical data refer mostly to the monopoles with a double magnetic
charge, $|Q_m|=2$ where the units are fixed by the Dirac 
quantization condition for the gluons. Classically, there are no stable solutions with
$|Q_m|=2$  \cite{bn} and, therefore, quantum effects seem to be absolutely
crucial even to introduce such monopoles. As a result, the theory
of these monopoles is in its infancy.

(b) There are recent
measurements of the interaction potential between the fundamental
monopoles with $|Q_m|=1$ on the lattice \cite{hoelbling},
which are introduced through the so called 't Hooft loop \cite{loop}. Unlike the case of 
the $|Q_m|=2$ monopoles the interaction of the fundamental 
monopoles is in fact quite well understood. The fact, 
which might be not well appreciated by the community.

(c) There exists surprisingly simple phenomenological description
of the properties of the $|Q_m|=2$ monopoles which are so poorly
understood on the purely theoretical side. We mean here models like
the Abelian Higgs model which provide 
quantitative support to the old idea of the dual-superconductor
mechanism \cite{confinement} and work surprisingly well at least in
some cases, for review and further references see \cite{review,baker}. 
  
In this mini-review we will emphasize some new points 
related to each of the items (a)-(c) listed above 
and which are based mostly on the 
original papers \cite{main,main1,main2}.
The new points, although they refer
to various topics, are unified by a common approach.
The starting point is that we consider monopoles within the
fundamental gluodynamics while the more traditional approach
is to introduce monopoles within an effective theory
intended to mimic QCD in the infrared region \cite{review,baker}.
Also, we understand the continuum gluodynamics rather as the limiting
case of the lattice formulation. As a result, one allows for certain 
singular gauge transformations which are not included
in more traditional frameworks. 

\subsection{Dirac monopole and Dirac string.}

The Dirac monopole, by definition, is associated with a radial
magnetic field similar to the electric field of a point-like charge,
$\bh= (4\pi)^{-1}\, Q_m \, (\br/r^3)$. One can easily construct a corresponding
vector potential:
\beq\label{dirac}
A_r~=~A_{\theta}~=0, ~~~A_{\phi}~=~\frac{Q_m}{4\pi}{(1+\cos \theta)\over r\sin\theta}.
\eeq
The analogy between the electric and magnetic charges is somewhat formal, 
however. Namely, because of the conservation of the magnetic flux, 
the radial magnetic field of the monopole should be supplemented
by the magnetic field of a string which brings in the flux spread
out uniformly by the radial component of the field.
Thus, we actually have
\beq
\bh~=~\bh_{rad}+\bh_{string}\,.
\eeq
 The presence of the string is exhibited,
in particular, by the explicit expression for 
the potential ${\bf A}$ above.

The Dirac string is unphysical and there is a number of constraints 
imposed on the
theory to ensure that the string does not produce any physical effect. 
First, there is the {\it Dirac veto} which forbids any direct interaction with the string.
The best known constraint is the {\it Dirac quantization condition}
which ensures the absence of the Aharonov-Bohm effect for the electrons
scattered on the string:
\beq
Q_e \oint_{string} {\bf A} \mathrm{d} {\bf x}~=~ Q_e\,Q_m ~=~ 2\pi k\,,
\label{quant}
\eeq
where $Q_e$ is the electric charge of the electron and $k$ is an integer number.
Let us also emphasize that naively the energy of the string is infinite in the ultraviolet:
\beq
\epsilon_{string}\sim \int (\bh_{string})^2 \, \mathrm{d}^3r \,\sim\,
{(Length)\over (Area)} \,\sim\, \Lambda_{UV}^2 (Length)\,,
\label{quadratic}
\eeq
where we used the fact that the magnetic flux is quantized (see above)
and that the cross section of the string denoted by $(Area)$ should tend
to zero at the end of the calculation.
Thus, we substituted $(Area)^{-1}$ by $\Lambda_{UV}^2$. 

The radial part of the magnetic field is also associated 
with an infinite energy:
\beq
\epsilon_{rad}~\sim\int (\bh_{rad})^2 \,\mathrm{d}^3 r~\sim~{1\over r_0}~\sim~\Lambda_{UV}\,.
\label{linear}
\eeq
Note that this ultraviolet divergence is linear, i.e. somewhat weaker than the
divergence due to the string, see Eq.~(\ref{quadratic}).

The infinite magnetic field of the string may have more subtle manifestations
as well. Consider interaction of two magnetic monopoles with magnetic charge $\pm Q_m$
placed at distance $R$ from each other. Then, by the analogy with
the the case of two electric charges, we would like to have
the following expression for the interaction energy:
\beq
\label{right}
\epsilon_{int}~=~\int \bh_{1,rad}\bh_{2,rad}  \,\mathrm{d}^3 r~=~
- \frac{Q_m^2}{4\pi}\,\frac{1}{R} \,.
\eeq
Note, however, that if we substitute the sum of the radial and string fields
for $\bh_{1,2}$, then we would have an extra term in the interaction energy:
\beq
\label{wrong}
\tilde{\epsilon}_{int}~=~
\int \left(\bh_{1,rad}\bh_{2,string}+\bh_{1,string}\bh_{2,rad}
\right)\,\mathrm{d}^3 r  ~=~
+ 2\,\frac{Q_m^2}{4\pi}\,\frac{1}{R}\,.
\eeq
In other words, the account of the string field would flip the
sign
of the interaction energy!
This contribution, although looks absolutely finite, is
of course a manifestation of the singular nature of the string
magnetic field, $|\bh_{string}|\sim (Flux)/(Area)$. Note that the
integral in (\ref{wrong}) does not depend on the shape of the string.

To maintain the unphysical nature of the Dirac string we should 
use a regularization scheme which would allow to get rid of these
singularities.  

\subsection{Lattice regularization.}

Since the monopoles are naively having divergent energy 
(or action) in the ultraviolet, the regularization is a crucial issue.
Moreover, we would like to follow the lattice formulation
since the monopoles are observed on the lattice.

Consider first the $U(1)$ case. As is emphasized in Ref. \cite{polyakov},
the lattice formulation implies that Dirac string
which produces no Aharonov-Bohm scattering costs no action as well.
The reason is very simple. The lattice action is written
originally in terms of the contour integrals like (\ref{quant})
rather than field strength $F_{\mu\nu}$:
\beq
S~=~ \sum\limits_{p} \,\mathrm{Re} \;\exp\{iQ_e \oint_{\diff p} A_{\mu}dx^{\mu} \}\,,
\eeq
where the sum is taken over all the plaquettes $p$. Thus, the condition (\ref{quant})
means absence of both the Aharonov-Bohm effect and the quadratic divergence 
(\ref{quadratic}) in the lattice regularization.
Moreover, it is straightforward to see that the interference term (\ref{wrong}) also vanishes.
Later, we will also discuss the case of the Dirac string which in the limit $g\to 0$
correspond to negative plaquettes in the lattice formulation. Its energy is infinite
in the continuum limit, in agreement with the naive estimate (\ref{quadratic}).
The interference term (\ref{wrong}), however, disappears in the lattice formulation
in this case as well.

Moreover, the lattice formulation 
naturally leads to the mo\-no\-po\-le--anti\-mo\-no\-po\-le potential (\ref{right})
without the unphysical string contribution (\ref{wrong}).

The radial field, $\bh_{rad}$ may also cause problems with infinite energy,
see (\ref{linear}). The lattice regularization is not much specific in that case,
however. The role of $r_0$ is simply played by the lattice spacing $a$.
Thus, the probability to find a monopole on the lattice is suppressed by
the action as:
\beq
\label{suppression}
e^{-S}~\sim~\exp (-const\cdot Q_e^{-2} \,L/a)\,,
\eeq
where $L$ is the length of the monopole trajectory, and the $Q_e^{-2}$
factor appears because of the Dirac quantization
condition (\ref{quant}) which relates the magnetic charge $Q_m$ to the inverse
electric charge.

Although the Eq.~(\ref{suppression}), at first sight, rules out monopoles as
physically significant excitations, the fate of the monopoles in the 
$U(1)$ case depends in fact on the value of the charge $Q_e$. The point is that
the entropy factor grows also exponentially with the length of the monopole 
trajectory:
\beq
(Entropy)~\sim~\exp(+const' \cdot L/a)\,,
\eeq
where the $const'$ is a pure geometric factor, not related to any
coupling constant like $Q_e$. As a result for $Q_e\sim 1$ there is a 
phase transition corresponding to the condensation of the monopoles. 
This phase transition, which is well studied on the lattice, is the first
and striking example of importance of the UV regularization
in the non-perturbative sector. Indeed, once the UV divergence
(\ref{quadratic}) is removed by the lattice regularization the monopoles
can modify the physics completely (for further comments see \cite{itep}).

\subsection{Classification of monopoles in non-Abelian theories.}

From now on, we will discuss monopoles in unbroken non-Abelian gauge theories,
having in mind primarily gluodynamics, i.e. quantum chromodynamics
without dynamical quarks. Moreover, for the sake of simplicity we will
consider only the $SU(2)$ gauge group.

A natural starting point to consider monopoles in non-Abelian theories
is their classification. There are actually a few approaches to
the monopole classification and it is important to realize both similarities
and differences between them.

{\it The dynamical, or $U(1)$ classification.}
Within this approach \cite{classification}, one looks for monopole-like
solutions of the classical Yang-Mills equations.
Where by the ``monopole-like'' solutions one understands
potentials which fall off as $1/r$ at large $r$, see Eq. (\ref{dirac}).
The basic finding is that there are no specific non-Abelian
solutions and all the monopoles can be viewed as Abelian-like embedded
into the $SU(2)$ group. Moreover, using the gauge invariance one can
always choose the corresponding $U(1)$ group as, say,
 the rotation group around the
third direction in the color space. According to this classification, 
the monopoles are characterized by their charge with respect to a $U(1)$
group and may have, therefore, charges,
\beq
|Q_m|~=~0, ~1,~2,~...~~~~. 
\eeq

{\it The topological, or $Z_2$ classification.}
The $Z_2$ classification \cite{tze} is based entirely on topological arguments.
Namely, independent types of monopoles
can be enumerated by considering the first homotopy group of the gauge group.
The SU(2) gauge group is trivial since $\pi_1(SU(2))=0$, while in the case of
the $SO(3)$, however,
\beq
\pi_1(SU(2)/Z_2)~=~Z_2\label{z2}
\eeq
and there exists a single non-trivial topological monopole. We will denote the magnetic charge
of such monopoles as $|Q_m|=1$. Note, however, that the charges
$Q_m=\pm 1$ are indistinguishable in fact. As for the charges $Q_m=2$
they are equivalent, from this point of view, to no magnetic charge
at all. 

The topological classification (\ref{z2}) is readily understood if one
tries to enumerate various types of the Dirac strings whose end points
represent monopoles under consideration.
Then there is only one non-trivial string, that is the one for which
Eq. (\ref{quant}) is satisfied for gluons but not for quarks.
Namely, because the $U(1)$ charge associated with gluons is twice as big as
that of the particles in the fundamental representation (quarks) 
we may have
\beq
\exp\{\, ig \oint A_{\mu}dx^{\mu} \,\} ~=~ -1
\eeq
and such a string is not visible for the isospin one particles.
On the other hand, the standard plaquette action is based on the phase factor evaluated
for particles in the fundamental representation. Which means, in turn, that
the Dirac string is piercing the negative plaquettes. This observation is the basis for
introducing the $|Q_m|=1$ monopoles via the 't~Hooft loop:
one changes the sign of $\beta$ ($\beta\equiv 4/g^2$) on a world sheet.
The boundary of this sheet corresponds to the end points of the Dirac
string, or the monopole trajectory.

\subsection{$Z_2$ monopoles.}

In principle, the $U(1)$ and $Z_2$ classifications are different.
Indeed, while the $U(1)$ classification allows for any integer
charge, the $Z_2$ classification leaves space only for a single non-trivial charge:
\beq
Q_m~=~0,1.
\eeq
The reconciliation 
of the two classifications is that the $U(1)$ solutions with 
$|Q_m|\ge 2$ are in fact unstable because of the presence massless charged
vector particles (gluons) \cite{bn}.
The instability of the solutions implies that even if 
the external sources with $|Q_m|\ge 2$ were introduced into the
vacuum state of the gluodynamics, charged gluons would fall onto 
the center because of the strong magnetic interactions. 
Moreover, one can imagine that as result of this instability
the charges fields $A^{\pm}$ are build up as well.

In a somewhat related way, one can demonstrate
the apparent irrelevance of the $|Q_m|=2$ monopoles 
by producing an explicit non-Abelian field configuration
which looks as a $|Q_m|=2$ monopole in its Abelian part
but has {\it no $SU(2)$ action} at all \cite{main}. This field 
configuration is a Dirac string with open ends,
which correspond to the monopole-anti-monopole pair separated by
the distance $R$.
In more detail, such a configuration is generated 
from the vacuum
by the following gauge rotation matrix:
\beq
\label{example}
\Omega~=~\left(\matrix{
e^{i\varphi}\sqrt{A_D} & \sqrt{1-A_D}\cr
-\sqrt{1-A_D}          &  e^{-i\varphi}\sqrt{A_D}\cr
}\right)\,,
\eeq
where $\varphi$ is the angle of rotation around the axis connecting
the monopoles  and $A_D$ is the $U(1)$ potential representing pure Abelian monopole pair:
\beq
A_{\mu}dx_{\mu}~=~
{1\over 2}\left({z_+\over r_+}-{z_-\over r_-}\right)d\varphi
~\equiv~ A_D(z,\rho)d\varphi\,,
\eeq
where $z_{\pm}=z\pm R/2$, $\rho^2=x^2+y^2$, $r_{\pm}^2=z_{\pm}^2+\rho^2$.
Note that the action associated with the Dirac string is considered in this
case zero, in accordance with  the lattice version of the theory
(for details see \cite{main}).

In this example, the monopoles with $|Q_m|=2$ are a kind of a pure
gauge field configurations carrying no action. 
Note that the Abelian flux is still transported along the Dirac string
and is still conserved for the radial field. 
What is lost, however, is the relation between the Abelian flux and action.
In the Abelian case non-vanishing flux means non-vanishing
magnetic field and non-vanishing action since the action density
is simply $\bh^2$. Now the action is $(F_{\mu\nu}^a)^2$ and the Abelian part
of the $F_{\mu\nu}^3$ can be canceled by the commutator term. 
This is exactly what happens in the example (\ref{example}) above.

It is somewhat more difficult to visualize dynamically the equivalence of
the $Q_m=\pm 1$ monopoles, also implied by the $Z_2$ classification.
The mechanism mixing the $Q_m=\pm 1$ solutions seems to be the following.
Imagine that we start with, say, $Q_m=+1$ solution. Then a Dirac
string carrying the flux corresponding to the $Q_m=-2$ can be superimposed
on this solution. It is important at this point that
such a Dirac string costs no action (or energy). Then the radial
magnetic field can also change its direction since it does not
contradict the flux conservation any longer. In a related language, 
one could say that the $|Q_m|=2$ monopoles are condensed in the
vacuum and that is why the magnetic charge can be changed freely
by two units. 

As far as interaction of two $|Q_m|=1$ monopoles is concerned, one might expect that
they would behave themselves as an monopole-antimonopole pair.
Indeed monopole and antimonopole would attract each other and thus represent the
lowest energy state of the system.

\subsection{Conclusions  \# 1}

Thus, the physics of the monopoles in the first approximation 
turns very simple. 

Namely, there exist only monopoles with $|Q_m|=1\equiv 2\pi/g$
where $g$ is the coupling constant of the non-Abelian $SU(2)$
theory.
The monopoles are infinitely heavy and can be introduced only as
external object through the 't Hooft loop. Their interaction is 
Abelian like:
\beq
V_{m\bar{m}}~=~-{Q_m^2\over 4\pi R}~=~-{\pi\over g^2 R}\,,
\label{potent}
\eeq
where $R$ is the separation between the monopoles. 

Clearly enough, this first, or classical approximation falls far beyond
an adequate description of the empirical data on the monopoles,
see the Introduction. Thus, we are invited to go into more advanced approaches 
which we would try to introduce step by step.

\section{Lagrangian approach.}

\subsection{The Zwanziger Lagrangian.}

There is a long standing interest 
in constructing the dual gluodynamics, for review and
further references see \cite{baker}.
The dual gluon, by definition, interacts with monopoles. 
The motivation is to realize in the field theoretical language
the dual superconductor model of
the quark confinement 
\cite{confinement} according to which the quarks are connected 
at large distances
by an Abrikosov-type vortex \cite{abrikosov}.
The key element is the construction of the non-Abelian monopoles, which are
usually modeled after the 't~Hooft--Polyakov solution. Namely, one introduces first
non-Abelian dual gluons interacting with Higgs fields and then 
assumes condensation of the Higgs fields which mimics the condensation of the
monopoles. In the realistic case of the $SU(3)$ gauge group one needs an octet
of dual gluons and three octets of the Higgs fields, all of them 
understood in terms of effective field theory valid in the infrared
region.

While such a construction might be viable as an effective theory,
we need in fact tools to describe interaction of non-Abelian
monopoles at arbitrary short distances as well \cite{main}. 
Indeed, in the lattice version of the theory
external monopoles can be introduced via the 't Hooft loop operator
\cite{loop} and in the continuum limit these monopoles are point like.
Thus, we are encouraged to consider the dual gluodynamics at short
distances, or at the {\it fundamental} level. 

It is natural to try a Lagrangian approach to the dual gluodynamics.
Indeed, in case of the same 't Hooft loop operator it is known that its expectation
value depends only on the boundary and not on
the shape of the Dirac string. Thus, it seems natural to introduce
a dual gluon which would interact 
directly with point-like monopoles.
In the context of electrodynamics, the idea is of course
very old and goes back to papers in Ref. \cite{zwanziger}.
There are successes and problems inherent to this approach,
for a review see \cite{blagoevich}. 

A well known example of Lagrangian which describes interaction
of a U(1) gauge fields with Abelian point-like monopoles is due to Zwanziger \cite{zwanziger}:
\beq
\label{Zw-action}
L_{Zw}(A,B)~=~ \frac{1}{2}(m\cdot[\diff\wedge A])^2 ~+~ \frac{1}{2}(m\cdot[\diff\wedge B])^2 ~+
\eeq
$$
+~\frac{i}{2}(m\cdot[\diff\wedge A])(m\cdot\dual{[\diff\wedge B]}) ~-~
\frac{i}{2}(m\cdot[\diff\wedge B])(m\cdot\dual{[\diff\wedge A]})
~+~ i\,j_e\cdot A ~+~ i\,j_m\cdot B\,,
$$
where $j_e,j_m$ are electric and magnetic 
currents, respectively, $m_{\mu}$ is a constant vector, $m^2=1$ and
$$
[A\wedge B]_{\mu\nu} = A_\mu B_\nu - A_\nu B_\mu\,, \qquad
(m \cdot [A\wedge B])_\mu = m_\nu [A\wedge B]_{\mu\nu}\,,
$$
$$
\dual{[A\wedge B]}_{\mu\nu} ~=~ \frac{1}{2}\,\varepsilon_{\mu\nu\lambda\rho}\,
[A\wedge B]_{\lambda\rho}\,.
$$
At first sight, we have introduced two different vector fields, $A,B$
to describe interaction with electric and magnetic charges,
respectively. If it were so, however, we would have solved a wrong problem because
we need to have a single photon interacting both with electric 
and magnetic charges. And this is what is achieved by the construct 
(\ref{Zw-action}). Indeed, the action (\ref{Zw-action}) is not diagonal
in the $A$, $B$ fields and one can convince oneself that
the form of the bilinear in $A,B$ interference terms in (\ref{Zw-action})
is such that the field strength tensors constructed on the potentials $A$ and $B$ are in fact related to each other:
\beq
F_{\mu\nu}(A)~=~\dual{F}_{\mu\nu}(B).
\eeq
Which means in turn that there are only two physical degrees of freedom
corresponding to the transverse photons which can be
described either in terms of the potential $A$ or $B$.  
Topological excitations, however, can be different in terms of $A$ and $B$.

The physical content of (\ref{Zw-action}) is revealed by the propagators
for the fields $A,B$.
In the $\alpha$-gauge one can derive:
\bea{c}
\label{propagators}
\langle A_{\mu} A_{\nu}\rangle ~=~
\langle B_{\mu} B_{\nu}\rangle ~=~
\frac{1}{k^2}\,(\delta_{\mu\nu}~-~(1-\alpha)\,\frac{k_\mu k_\nu}{k^2})\,,
\\
\\
\langle A_{\mu} B_{\nu}\rangle ~=~
- \langle B_{\mu} A_{\nu}\rangle ~=~
\frac{i}{k^2 (km)} \; \dual{[m\wedge k]}_{\mu\nu}\,.
\eea
The propagators should reproduce, as usual the classical solutions. And indeed,
the $\langle AA \rangle$, $\langle BB \rangle$ propagators describe
the Coulomb-like interaction of two charges and magnetic monopoles, respectively.
While the $\langle AB \rangle$ propagator reproduces interaction of the magnetic field
of a monopole with a moving electric charge.The appearance of the poles in $(k\cdot m)$
is a manifestation of the Dirac strings.

To summarize, the Zwanziger Lagrangian in electrodynamics \cite{zwanziger} reproduces
the classical interaction of monopoles and charges. Upon the quantization,
it describes the correct number of the degrees of freedom associated with the photon.

\subsection{Dual gluon as an Abelian vector field.}

Now, if we would approach the problem of constructing a Zwanziger-type Lagrangian 
for the dual gluodynamics, we immediately come to a paradoxical conclusion that the
dual field, if any, is Abelian. Indeed, monopoles associated with, say, $SU(N)$ gauge group
are classified according to $U(1)^{N-1}$ subgroups \cite{classification}
and might be realized as a pure Abelian objects. Thus, there is no place for a
non-Abelian dual gluon because the monopoles do not constitute representations
of the non-Abelian group.

The function of the classical Lagrangian is, first of all,
to reproduce the classical interactions of the monopoles  
and charges. It is rather obvious that the potential (\ref{potent})
can be derived in the classical approximation from the Lagrangian:
\beq
\label{b-action}
L_{dual}(A^a,B) ~=~ \frac{1}{4} (F^a_{\mu\nu})^2 ~+~
\frac{1}{2} ( m\cdot [\,\diff\wedge B - i \,\dual{G}\,] )^2
~+~ i\,j_m B ~+~ i\,j^a_e A^a\,,
\eeq
where $a=1,2,3$ is the color index,  $j_m$ is the magnetic current and 
$F^a_{\mu\nu}$ is the non-Abelian field strength tensor. The Lagrangian (\ref{b-action})
also contains vector field $n^a$, $n^2=1$ in the adjoint representation and antisymmetric
tensor $G_{\mu\nu}$ is the 't~Hooft tensor \cite{monopole}:
\beq
\label{tensor}
G_{\mu\nu} ~=~ n^a F^a_{\mu\nu}~-~\varepsilon^{abc}\, n^a \, (D_{\mu}n)^b \, (D_{\nu}n)^c \,.
\eeq
Let us add a few comments on the meaning and rules of using the Lagrangian (\ref{b-action}).

(a) First, if the magnetic current is vanishing, $j_m=0$ then 
the integration over the field $B$ reproduces
the standard Lagrangian of the gluodynamics.

(b) As far as the quantization is concerned, the Lagrangian (\ref{b-action})
reproduces the correct degrees of freedom of the free gluons. Indeed,
in the limit $g\to 0$ and for $n^a=\delta^{a,3}$ the Lagrangian (\ref{b-action}) 
becomes:
$$
L_{dual}(A^a,B) ~=~ \frac{1}{4}(\diff\wedge A^1)^2 + \frac{1}{4}(\diff\wedge A^2)^2 +
i\,j_m B + i\,j^a_e A^a + 
$$
\beq
+ \frac{1}{2}[m\cdot(\diff\wedge A^3)]^2 + \frac{1}{2}[m\cdot(\diff\wedge B)]^2 +
\eeq
$$
+\frac{i}{2} [m\cdot(\diff\wedge A^3)] [m\cdot\dual{(\diff\wedge B)}] - 
\frac{i}{2} [m\cdot\dual{(\diff\wedge A^3)}] [m\cdot(\diff\wedge B)]\,,
$$
which is essentially the Zwanziger Lagrangian (\ref{Zw-action}). Quantization at this point
is the same as in the case of a single photon.

(c) Already in the Zwanziger example (\ref{Zw-action}) we have seen
that the fields that are mixed up in the Lagrangian have
a common source. Namely, in case of the electrodynamics
$\diff\, \dual{F} (A)=\diff F(B)=j_m$. Since 
it is known \cite{monopole} that the monopoles in non-Abelian theories
serve as source for the 't Hooft tensor (\ref{tensor}) one expect
from the very beginning that in case of the gluodynamics
the (dual) field strength tensor build up on the dual gluon field $B$
is mixed up with the 't Hooft tensor constructed in terms of the gluon filed $A$.
And, indeed, this is true for (\ref{b-action}).

(d) The emergence of the vector $n^a$ is of crucial importance in the
Lagrangian (\ref{b-action}). The point is that the origin of the vector $n^a$
goes back to choosing the color orientation of the monopoles.
As is emphasized above the monopole solutions are Abelian in nature
which means, in particular, that they can be rotated
to any direction in the color space by gauge transformations.
Thus, picking up a particular $n^a$ is nothing else but using
the gauge fixing freedom. Therefore, we can either average over the directions of $n^a$ or fix $n^a$ 
but evaluate only gauge invariant quantities, like the Wilson loop (note somewhat 
similar remarks in Ref. \cite{korthals}).

(e) The $Z_2$ nature of the monopoles is manifested in the freedom
of changing $n^a\to -n^a$, $B_\mu \to -B_\mu$. Indeed, under such transformation
the monopole with the charge 
$Q_m=+1$ is transformed into a monopole with $Q_m=-1$
and vice versa. In the language we used above such a transformation
corresponds to adding a Dirac string with a double magnetic flux.
We see that the averaging over $\pm n^a$ is a part of the overall
averaging over all possible embedding of the $U(1)$ into the $SU(2)$ gauge group.

An apparent application of (\ref{b-action}) would be evaluating the running of the
coupling $g$ in the expression (\ref{potent}). And, indeed, 
exploiting the Lagrangian (\ref{b-action}) one can approach the problem 
of the running of the coupling in a way similar to the case 
of pure electrodynamics, for a review and further references see 
\cite{blagoevich}. We comment on this approach below.

\subsection{Radiative corrections.}  

We will consider now the radiative corrections to the Coulomb-like
interaction  (\ref{potent}) at short distances. Obviously enough, 
one would expect
that the radiative corrections result in the standard, non-Abelian
running of the coupling $g^2$.
Which is indeed our main conclusion. Moreover, since for a 
constant vector $n^a$ the non-Abelian monopole essentially
coincides with the Dirac monopole, there is no much specific about
the derivation of the running of the coupling. And, indeed, our 
considerations overlap to a great extent with those given in 
the original papers \cite{calucci,goebel} and in the reviews 
\cite{coleman,blagoevich}. Still,we feel that it is useful to present
the arguments, may be in a new sequence, to emphasize the points crucial for
our purposes. 

Let us emphasize from the very beginning that the evaluation of the radiative
corrections addresses in fact two different, although closely related
problems. That is, running of the coupling and stability of the classical
solutions. Both aspects are unified, of course, into evaluation
of a single loop in the classical background.
However, the running of the coupling can be clarified by
keeping track of the ultraviolet logs, $\ln\Lambda_{UV}$ alone 
and are universal since in the ultraviolet all the 
external fields can be neglected. 
Therefore, the coefficient in front of $\ln \Lambda_{UV}$
can be found by evaluating the loop graph with two external
legs, i.e. the graph corresponding to the standard
polarization operator in perturbation theory.
This is true despite of the fact that the 
monopole field is strong 
(i.e. the product of the magnetic and electric coupling
is of order unity). On the other hand the stability
of the classical solution is decided by the physics in the infrared.
Here one needs to consider the particular dynamical system, monopoles
in our case, and the fact that the magnetic charge is of order $1/g$
can be crucial. 

Consider first the running of the coupling. 
Moreover, for the sake of definiteness we concentrate on the
Dirac monopole with the
minimal magnetic charge interacting with electrons
and in one-loop approximation
\cite{blagoevich,eg,calucci,goebel}. The crucial point here is that
only loops with insertion of two external (i.e., monopole) fields can
be considered despite of the fact that there is no perturbative
expansion at all. Indeed, considering more insertions make the
graphs infrared sensitive, with no possibility for $\ln\Lambda_{UV}$
to emerge.

Then, the evaluation of, say, first radiative correction to  the propagator
$\langle B_{\mu}B_{\nu}\rangle$ in the Zwanziger formalism  (\ref{propagators})
seems very straightforward and reduces to taking a product of two
$\langle AB \rangle$ propagators and inserting in between the standard polarization operator 
of two electromagnetic currents. 
The result is \cite{eg}:
\beq
\langle B_{\mu}B_{\nu}\rangle(k)
~=~{\delta_{\mu\nu}\over k^2}(1-L)
+{1\over (k\cdot m)^2}(\delta_{\mu\nu}-m_{\mu}m_{\nu}) L,\label{sch}
\eeq
$$
L~=~{\alpha_{el}\over 6}\ln{\Lambda_{UV}^2/k^2}
 $$
and we neglect the electron masses so that the infrared cut-off
is provided, in the logarithmic approximation, by the momentum $k$.
 
At first sight, there is nothing disturbing about the result (\ref{sch}).
Indeed, we have a renormalization of the original propagator which
is to be absorbed into the running coupling,
and a new structure with the factor $(k\cdot m)^{-2}$ which
is non-vanishing, however, only on the Dirac string. 
The latter term would correspond renormalization of the Dirac-string 
self-energy which we do not follow in any case 
since it is included into self-energy of the external monopoles.
What is, actually, disturbing is that according (\ref{sch})
the magnetic coupling would run exactly the same as the electric charge,
$$
\langle A_{\mu}A_{\nu}\rangle(k) ~=~ (1-L){\delta_{\mu\nu}\over k^2}\,,
$$
violating the Dirac quantization condition.

The origin of the trouble is not difficult to figure out. Indeed, using the propagator 
$\langle AB \rangle$ while evaluating the radiative corrections is equivalent, of course,
to using the full potential corresponding to the Dirac monopole $A_D^{cl}$. Then, switching on the
interaction with electrons would bring terms like  $A_D^{cl}\bar{\psi} \gamma\psi$. Since $A_D$ 
includes the potential of the string electrons do  interact with the Dirac string and we are
violating the Dirac ``veto'' which forbids any direct interaction with the string.  

Let us demonstrate that, indeed, the incorrect treatment of the Dirac string
changes the sign of the radiative correction.
This can be done in fact in an amusingly simple way.
First, let us note that it is much simpler to remove the string
if one works in terms of the field strength tensor, not the potential.
Indeed, we have $\bh~=~\bh_{rad}+\bh_{string}$
while in terms of the potential ${\bf A}$ any separation of the
string would be ambiguous (see Eq.~(\ref{dirac})).

Thus, we start with relating the potential, or energy to the
interference term in the $\bh^2$ field:
\beq
V_{m\bar{m}}~=~ \int \bh_1\cdot\bh_2 \, \mathrm{d}^3 r\,.
\eeq
Now, it is not absolutely trivial, how we should understand
the product $\bh_1\cdot \bh_2$. Indeed, we emphasized in section 1.2
that the string field is to be removed from this interference term,
see Eq (\ref{wrong}).
Thus, in the zero, or classical approximation we have:
\beq
\bh_1\cdot\bh_2~\equiv~\bh_{1,rad}\cdot\bh_{2,rad}\,.
\eeq
However, if we use the standard technique of an external field:
\beq
A_{\mu}~=~A_{\mu}^{class}+a_{\mu}
\eeq
and substitute (\ref{dirac}) as the classical background then
the first radiative correction would bring the product of
the total $\bh_1\cdot\bh_2$ which includes also the string
contribution\footnote{At this point we assume in fact that 
$\Lambda_{UV}$ is smaller than the inverse size of the string,
which is convenient for our purposes here.
Other limiting procedures could be considered as well,
however.}. Indeed, the result in the log approximation
would be as follows:
\beq
\delta (\bh_1\cdot\bh_2)~=~L(\bh_{1,string}+\bh_{1,rad})\cdot
(\bh_{2,string}+\bh_{2,rad})~=~-L\bh_{1,rad}\cdot\bh_{2,rad},\label{delta}
\eeq
where at the last step we have used the observation (\ref{wrong}).

Now, it is clear how we could ameliorate the situation.
Namely, to keep the Dirac string unphysical we should remove
the string field from the expression (\ref{delta})
which arises automatically if we use the propagators (\ref{propagators})
following from the Zwanziger Lagrangian.
Thus, we introduce:
\beq
(\bh_1\cdot \bh_2)^{'}~\equiv~\bh_{1,rad}\cdot\bh_{2,rad}\label{corrected}
\eeq
and change $\bh_1\cdot\bh_2$ in the expression (\ref{delta})
into (\ref{corrected}) so to say by hand. The justification is that
we should remove the effect of the string field from any observable.

Then we reverse the sign of the radiative correction and the final
result is
\beq
(V_{m\bar{m}})_{class}~\equiv~-{\pi\over g_0^2}{1\over R}~
\to~-{\pi\over g^2(R)}{1\over R}\,.
\eeq

One might wonder, how it happens that the couplings
in the electric and magnetic potential run in opposite ways.
Indeed, now we reduced the product $\bh_1\cdot\bh_2$ to exactly the same
form as the product ${\bf E}_1\cdot {\bf E}_2$ in case
of two electric charges (since the radial magnetic and electric 
fields are
the same, up to a change of the overall constants).
The resolution of the paradox is that the renormalization 
of the electric and magnetic fields are indeed similar 
in the language of the Lagrangian. However, the small 
corrections to the Lagrangian and Hamiltonian are related as:
\beq\label{delta1}
\delta L~=~-\delta H.
\eeq
Since ${\bf E}^2$ and ${\bf H}^2$ enter with the same sign into the
expression for the Hamiltonian and with the opposite signs into
the Lagrangian, Eq
(\ref{delta1}) implies that the running of the couplings
in the electric $V_{e}$ and magnetic $V_m$ potentials are opposite in sign.  
Which is, of course, in full agreement with expectations since 
$V_e\sim g^2$ and $V_m\sim g^{-2}$.

Thus, it is not difficult to derive the running of the magnetic coupling
following only the ultraviolet log, $\ln\Lambda_{UV}$.
Note, however, that the same arguments would go through without
change if we started with, say, monopoles with $Q_m=2$.
But such monopoles are unstable \cite{bn} and this is a much
more drastic effect than the would-be running of the coupling.
There are also more subtle mechanisms which can be brought in
by radiative corrections. In case of the same Dirac monopole interacting with
electrons \cite{yang} consideration of the modes reveals that the
Hamiltonian is in fact non Hermitian. As a result the classical field
approximation is not adequate and one should consider the corresponding
field theory, or the monopole catalysis \cite{callan}.

Thus, to investigate the stability of the classical soluition
one has, generally speaking, to consider all orders in
$g_e\, g_m\sim 1$. It is known that single monopoles with $Q_m=1$ are stable.
The stability of the monopole-antimonopole system, which we are interested
in, has never been investigated analytically in detail because
of the complexity of the problem. However, there is no known mechanism which
could cause instability of the classical monopole-antimonopole solution.
Moreover, we checked numerically that the classical solution is
indeed stable \cite{main}.

\subsection{Why the ``right way'' is correct?}

Thus, our exercise with evaluating the running of the magnetic 
coupling has brought mixed results. On one hand, we were able to
derive that the product of electric and magnetic coupling constants
is not renormalized, as one would expect.
On the other hand, to derive this we had to go actually beyond the Lagrangian
approach and remove the effect of the magnetic field of the string.
Now we ask the next question, why this removing was the correct procedure.

Let us reexamine the grounds for the Lagrangian approach, in their
generality. Any monopole involves also a Dirac string and, as a result,
a world sheet, not just particle trajectories.
If we stop here, then the conclusion would be that there is no Lagrangian
approach to the problem. 
However, we are aware that the 't Hooft loop operator depends 
only on its boundary, which is the monopoles trajectory, $j_m$.
And this is the real basis for the hopes for the Lagrangian formulation.
Now, we see that the Dirac veto is not respected by the Lagrangian 
formulation and, therefore, the possibility arise that the world sheet
swept by the Dirac string is still somehow important.
Thus, we will outline in this subsection an approach \cite{main}
which is based on derivation of a continuum 
analog of the lattice 't~Hooft loop operator and avoids any direct use of Lagrangians.

The general one-plaquette action of $SU(2)$ lattice gauge theory (LGT) 
can be represented as:
\beq
S_{lat}(U)~=~{4\over g^2} \,\sum\limits_p \, S_p\left(1-{1\over 2}TrU[\partial
p]\right),\label{plaquette}
\eeq
where $g$ is the bare coupling, $\partial p$ is 
the boundary of an elementary plaquette $p$,
the sum is taken over all $p$, $U[\partial p]$ is the ordered 
product of link variables $U_l$ along $\partial p$. 
In particular, if $S_P(x)=x$ then (\ref{plaquette})
is the standard Wilson action.
The exponent of the lattice field strength tensor $F_p$ is defined in terms of
$U[\partial p]$:
\beq
U[\partial p]~=~e^{i\hat{F}_p}~=~
\cos\big[{1\over 2}|F_p|\big]+i\tau^an^a\sin\big[{1\over 2}|F_p|\big],\label{partial}
\eeq
where $\hat{F}=F^a\cdot \tau^a/2, |F|=\sqrt{F^aF^a}$
and we define $n^a_p~=~F_p^a/|F_p|$ for
$|F_p|\neq 0$, while $n^a_p$ is an arbitrary unit vector for $|F_p|=0$.

The lattice action (\ref{plaquette}) 
depends only on $\cos\big[{1\over 2}|F_p|\big]$. Therefore the action
of the $SU(2)$ LGT possesses not only the usual gauge symmetry, but allows
also for the gauge transformations which shift the field strength tensor
by $4\pi k$, $|F_p|\to |F_p|+4\pi k$,
$k\in Z$:
\beq
e^{i\hat{F}_p}~=~\exp\{i|F_p|\hat{n}_p\}
~=~\exp\{i(|F_p|+4\pi)\hat{n}_p\}~=~
\exp\{i(F_p^a+4\pi n^a_p)\tau^a/2\}.
\eeq

Thus, the symmetry inherent to the lattice formulation can be represented as:
\beq
F_p^a~\to~F_p^a+4\pi n_p^a,~~~\vec{F}_p\times \vec{n}_p~=~0,~~~n_p^2=1.
\label{symmetry}\eeq
The symmetry (\ref{symmetry}) is absent in the conventional continuum
limit, $\int(F_{\mu\nu}^a)^2d^4x$. 
Note that in the continuum limit $n_p^a$ becomes a singular two-dimensional
structure  $\dual{\Sigma}^a_{\mu\nu}$ which is representing the Dirac string world sheet.

So far we discussed an invisible Dirac string, which is 
nothing else but a generalized (or singular) gauge transformation.
The Dirac string corresponding to the fundamental monopole corresponds
to the phase factor $-1$ and we can obtain, therefore, an expression for
a continuum analog of the 't Hooft loop by substituting:
\beq
F_{\mu\nu}^a~\to~F^a_{\mu\nu}+2\pi\dual{\Sigma}^a_{\mu\nu}.
\eeq  
In this way we come to the following definition of the 't Hooft loop operator
in the continuum:
\beq
\label{tHooft-general}
H(\Sigma_\cC) =
\exp\left\{
{1\over 4g^2}\int d^4 x \; \left[
\left(F^a_{\mu\nu}\right)^2
-
\left(F^a_{\mu\nu} + 2 \pi \dual{\Sigma}^a_{\cC \; \mu\nu}\right)^2
\right]\right\}\,,
\eeq
\beq
\label{Sigma-colored}
\Sigma^a_{\cC\;\mu\nu} = \int d^2\sigma_{\mu\nu} \; n^a(\sigma) \; 
\delta^{(4)}(x-\tilde{x}(\sigma))\,,
\eeq
where the surface $\Sigma^a_\cC$ spanned on the contour $\cC$ is assumed to be non-intersecting. 
The unit three-dimensional vector field $n^a(\sigma)$, $\vec{n}^2=1$ 
is defined on the 
world-sheet:
\beq
\label{tn}
n^a(\sigma) ~=~ (t \cdot \dual{F}^a ) \; \left[{(t \cdot \dual{F}^b )^2}\right]^{-1/2}\,,
\qquad
(t \cdot F^a) = t_{\mu\nu}(\sigma) \; F^a_{\mu\nu}(\tilde{x})\,,
\eeq
\beq
\label{tg}
t_{\mu\nu}(\sigma) =  {1\over \sqrt{g}} \;
\varepsilon^{\alpha\beta} \; \diff_\alpha \tilde{x}_\mu \; \diff_\beta \tilde{x}_\nu\,,
\qquad
t^2_{\mu\nu} = 2\,,
\qquad
g(\sigma) = \mathrm{Det}[\; \diff_\alpha \tilde{x}_\mu \; \diff_\beta \tilde{x}_\mu \;]\,.
\eeq
Therefore $n^a(\sigma)$ is not an independent variable, it is completely determined by 
the components of the field strength tensor $F^a_{\mu\nu}$. On the set of points
where $(t \cdot \dual{F}^a) = 0$ the direction of $n^a(\sigma)$ is arbitrary.
It can be shown \cite{main1} that the Eq.~(\ref{tHooft-general})-(\ref{tg}) define the correct
't~Hooft loop operator the expectation value of which depends only on the contour $\cC$,
not on the particular position of the surface $\Sigma_\cC$.

Consider now the equations of motion in presence of the 't~Hooft loop operator:
\beq
D_{\nu}\left( F_{\mu\nu}(A)+2\pi\dual{\Sigma}_{\mu\nu}\right)~=~0,\label{em}
\label{surface}
\eeq
which should be supplemented by the Bianchi identities:
\beq
D_{\nu} \dual{F}_{\mu\nu}~=~0.
\eeq
To appreciate the meaning of the equation of motion (\ref{em})
let us choose the gauge such that $\Sigma_{\mu\nu}^a$ has a constant color orientation 
characterized by the vector $n^a_0$.
A particular solution of (\ref{surface}) may be found within the anzatz $A^a_\mu = n^a_0 \, A_\mu$,
for which Eq.~(\ref{surface}) reduces to:
\beq
\diff_{\nu}\left(\diff_{[\mu}A_{\nu]}\right)~=~-2\pi\diff_{\nu}
\dual{\Sigma}_{\mu\nu}.\label{cl.eq}
\eeq
The solution of this equation in the Landau gauge,
\beq
A_{\mu}^a~=~-n^a_0\cdot 2\pi {1\over \Delta}\diff_{\nu} \dual{\Sigma}_{\mu\nu},
\eeq
corresponds to the gauge potential of an Abelian monopoles current
$\partial\Sigma$ embedded into the $SU(2)$ group.
Thus, $\Sigma_{\mu\nu}$ is the Dirac world sheet.

Derivation of the classical equations of motion (\ref{cl.eq})
is the first step in deriving the interaction of
the fundamental monopoles outside any Lagrangian framework.

One could consider along these lines also the radiative corrections
\cite{main2}. We will not go into details here but let us mention, 
how it comes about that the ``Dirac veto'' is observed
and virtual particles do not interact with the Dirac string.
We will substantiate this point here on the example of
the spin interaction. Since the Yang-Mills quanta possess
spin, there exists interaction which is generalization of
the non-relativistic expression ${\bf \sigma\cdot H}$. In particular, if there exists classical
field directed in third direction in the color space, $(F_{\mu\nu}^3)_{cl}$ then its interaction
with the quantum charged fields $a^{\pm}_{\mu}$ contains the term
\beq
g\,(F_{\mu\nu}^3)_{cl}\,a^+_{\mu}a^-_{\nu}\,.
\label{classical}
\eeq
In the Zwanziger formalism, the $(F_{\mu\nu}^3)_{cl}$ means the whole magnetic field,
the field of the string including. Then  the interaction (\ref{classical})
brings the term ${\bf H}_{1,string}\cdot{\bf H}_{2,radial}$ on the level
of the quantum corrections. As we emphasized in sect.~1.2, this term
actually proportional to ${\bf H}_{1,radial}\cdot{\bf H}_{2,radial}$
which is responsible for the coupling running. In this way
the term ${\bf H}_{1,string}\cdot{\bf H}_{2,radial}$, if it arises,
brings in a ``wrong'' sign of the radiative correction.

On the other hand, in our formulation of the continuum analog 
of the 't Hooft loop operator, see Eq. (\ref{tHooft-general}),   
there is no spin interaction of the virtual particles with the string magnetic field.
This is crucial to prove \cite{main2} that the coupling governing
the monopole-antimonopole interaction indeed runs as $g^{-2}$.

\subsection{Conclusions \# 2}

We considered in fact two different points. First we argued that
the dual gluon is a $U(1)$ gauge boson. The $SU(2)$ invariance is to be maintained
either by integrating over all the possible embedding of the (dual) $U(1)$
into $SU(2)$ or by constraining calculations to gauge invariant quantities,
like the Wilson loops.

Second, we discussed how far one can go with a Lagrangian formulation
of the dual gluodynamics a la Zwanziger. To test the Lagrangian
approach we evaluated the running of the coupling in
the monopole-antimonopole potential. The conclusion is that
one can get the correct running of the coupling  by imposing
the Dirac veto which forbids the interaction of virtual particles
with the Dirac string. This requirement is not inherent to the Lagrangian approach
(the same is true for the Zwanziger Lagrangian in the $U(1)$ case), however.
It can be derived by studying the continuum analog of the 't Hooft loop
operator.

\section{Monopoles with $Q_m$=2.}
\subsection{$Q_m=2$ monopoles as quantum objects.}

So far we discussed the fundamental monopoles $|Q_m|=1$ which can
be visualized as classical infinitely heavy objects. Because of
infinite mass, they can be used only as probes of the QCD vacuum but play
no dynamical role by themselves. The monopoles with the double charge
$|Q_m|=2$ are very different. As discussed above, they do not exist
on the classical level. On the other hand, there exist very simple
arguments that they can play dynamical role on the quantum
level. As far as the fundamental Lagrangian is concerned, the only role
of the quantum corrections is the running of the non-Abelian coupling.
In particular, if we consider a lattice coarse enough then $g_{SU(2)}$ becomes of order unity.
Obviously, the same coupling governs the physics associated with any 
$U(1)$ subgroup of the $SU(2)$. However, if the coupling $g_{U(1)}$ becomes
of order unity, then there is a phase transition associated with the monopole
condensation \cite{polyakov}. Thus, one can argue that the running will be
stopped by the monopole condensation, if not by something else already
at smaller values of $g_{SU(2)}$.

Thus, it is very natural to assume that the monopole condensation occurs also in 
QCD since the running of the coupling allows to scan the physics at all
the values of $g_{SU(2)}$ until one runs into a phase transition.

However, even if one accepts such speculations, there remains a very important
unresolved question. Namely, it is not clear which $U(1)$ subgroup of
the full non-Abelian group is to be selected
as the classification group for the monopoles.
The most common approach here is to rely on the empirical data.
In a way, it is forced on us since the phase transition
is expected to happen at $g^2_{SU(2)}\sim 1$ where
analytical approaches are hardly possible.
{}From the lattice simulations it is knwon that the monopoles
in the Maximal Abelian projection appear to be most relevant, 
see \cite{review} for review and further references.  

Instead of reviewing this material once more -- which would take us far beyond the scope
of the present article -- we will highlight some features of new kind of monopoles
introduced in Ref. \cite{main1}. The basic idea behind this construction is to make
monopoles as much geometrical objects as possible. 

\subsection{``Geometrical'' monopoles.}

The construction of the new kind monopoles is in few steps which
we will briefly outline  now.

({\it i}) The usual starting point to introduce monopoles is 
to fix some $U(1)$ for the whole lattice and 
then look for the Dirac strings and monopoles with respect to this $U(1)$.
The starting point of \cite{main1} is somewhat different.
Namely, it is the observation that each Wilson loop
defines in a natural way its own $U(1)$.
Indeed, turn back to the expression (\ref{partial})
for the plaquette action which is actually true for any Wilson loop.
Then, it is clear that each Wilson loop defines the vector
$\hat{F}_p$ and the ``natural'' $U(1)$ is the group of rotation 
around this vector (in the color space). In this way, one can define
a $U(1)$ group for each plaquette. The definition of the $U(1)$ subgroup
varies from one plaquette to another, emphasizing the non-Abelian nature
of the underlying theory.

({\it ii}) The plaquette action is $1/2\cos\phi$ and is invariant under
$\phi_p\rightarrow \phi_p+2\pi k$, as is emphasized in sect.~1.3.
Now to detect the Dirac strings we should be able to
somehow define the integer $k$. For a plaquette, the natural decomposition is 
\beq
\phi_p~=~\phi_1+\phi_2+\phi_3+\phi_4,\label{decomposition}
\eeq
where the phases $\phi_i~(i=1,...,4)$ are associated with the
corresponding links. 
The decomposition (\ref{decomposition}) comes about naturally in the basis
of the {\it coherent states}. Indeed, for a particular coherent state
the whole evolution may be reduced to a phase factor:
\beq
|\psi(t)\rangle~=~e^{i\phi(t)}|\psi(0)\rangle.
\eeq
Moreover, the coherent states can be explicitly constructed in terms of the link matrices,
for details and further references see \cite{main1}.
As a result, for any given lattice fields configuration, one can determine
$k$ and detect the Dirac strings in this way. The monopoles are defined then as 
the end points of the strings.

({\it iii}) The phase $\phi (t)$ can in fact be decomposed into the 
dynamical and Berry phase. It is useful for this purpose to introduce
a single-valued state vector $|\tilde{\psi}\rangle$ defined as
\beq
|\tilde{\psi}(T)\rangle~=~|\tilde{\psi}(0)\rangle\,,
\eeq
where $T$ is the period of the motion so that at $t=T$ the system
comes back to the same point in the parameter space as at the moment $t=0$.
Then 
\beq
\phi (T)~=~\delta+\gamma~=~-\int_{0}^{T}\langle\tilde{\psi}|H|\tilde{\psi}\rangle
+i\int_{C}\langle\tilde{\psi}|{\partial\over \partial \lambda_i}
|\tilde{\psi}\rangle \, d\lambda^i\,,
\eeq
where $\lambda_i$ are parameters, $\lambda_i(T)=\lambda_i(0)$ and
$C$ is a closed contour in the parameter space.

\subsection{Choice of the gauge and the numerical results.}

The steps ({\it i}) -- ({\it iii}) described above fully determine
monopoles
as geometrical objects. As a mathematical construct, it certainly appears
very appealing. However, from the physical point of view the
crucial observation is the gauge dependence of the monopoles
constructed in this way. As a result, the monopoles are devoid, generally
speaking of any physical meaning. It is amusing that one can actually specify the conditions for
the monopoles to be physical objects. In particular, the monopole density $\rho$
should satisfy the renormgroup equation:
\beq 
\rho~=~\mathrm{const}\;\cdot\;\beta^{153/121}\, \exp\left( -{9\pi^2\over 11}\beta \right)\label{rg},
\eeq
where $\beta\equiv 4/g^2$. 
The condition (\ref{rg}) is a very strong constraint and there is no much surprise that
the monopoles defined according to the procedure outlined in the
preceding subsection, generally speaking, do not satisfy (\ref{rg}).

To continue with the physics, we need a physically motivated 
choice of the gauge. At first sight, such a choice is impossible.
However, one can argue \cite{main1,stodolsky}
that the Lorenz gauge is a proper gauge. The Lorenz gauge on the lattice
is defined by the requirement that the functional
\beq
R~=~\sum\limits_l \left(1-\frac{1}{2} \mathrm{Tr} U_l \right)\label{lorenz}
\eeq
is minimal on the gauge orbit ($U_l$ denote the link matrices).
In the naive continuum limit (\ref{lorenz}) reduces to $R~=~1/4\int(A_{\mu}^a)^2$.

The logic behind the choice (\ref{lorenz}) is as follows. In the continuum
limit both the Dirac strings and monopoles correspond to singular gauge potentials
$A_{\mu}^a$. It is easy to imagine, therefore, that one can generate an arbitrary number
of spurious strings and monopoles by going to arbitrary large potentials $A$,
so to say inflated by the gauge transformations. On the other hand, by minimizing
potentials one may hope to squeeze the number
of the topological defects to its minimum and these topological defects
may be physically significant.

And, indeed, the numerical simulations indicate that the geometrical monopoles
defined in the Lorenz gauge are physical objects, i.e. their density
satisfies the condition (\ref{rg}). There are also other indications that
the geometrical monopoles are physical \cite{main1}. 
For example, there is an excess of the non-Abelian
action associated with them.

\subsection{Conclusions \# 3.}

Monopoles with $Q_m=2$ unify properties of field-theoretical and statistical objects.
Namely, on one hand the monopoles are defined locally in terms of the link
matrices. However, the link matrices are gauge dependent and existence
or non-existence of a monopole at a particular point is devoid of physical
meaning for this reason. On the other hand, if one introduces gauge
fixing in a physically reasonable way, the statistical properties of the monopoles,
such as their density, satisfy very non-trivial renormgroup constraints
and demonstrate their physical significance.

\section{Phenomenological applications.}

\subsection{The effective Lagrangian and the Casimir scaling.}

The standard way to develop a phenomenology is to assume that the monopoles
condense. We will follow the suit and modify the Zwanziger Lagrangian
(\ref{b-action}) by adding the effective Higgs interaction where the role
of the Higgs field is played by the monopole field $\phi_m$:
\beq
S_{eff}~=~S_{dual}(A^a,B)+S_{Higgs}(B,\phi_m)\label{higgs}\,,
\eeq
where $S_{Higgs}$ is the standard action of the Abelian Higgs model.
The vacuum expectation value of the Higgs, or monopole field is, of course,
of order $\Lambda_{QCD}$.

Despite its apparent simplicity, Eq. (\ref{higgs}) is highly speculative.
Namely, it unifies so to say fundamental gluons, $A^a$, their dual counterpart
$B$ which is an Abelian gauge boson, and $\phi_m$ which is presumably an effective
scalar field. One may justify the use of (\ref{higgs}) by assuming that the effective
size of the monopoles with $Q_m=2$ is in fact numerically small, although
generically it is of order $\Lambda_{QCD}$.
While in our presentation here we follow mostly the lines of Refs.~\cite{main,main2,ss},
let us note that similar consequences arise within the models \cite{shuryak,schmidt}
also introducing a new mass scale.

What is also specific about the Lagrangian (\ref{higgs}) is that the dual gluon is a $U(1)$ gauge boson.
The color symmetry is maintained by averaging over all possible embeddings of the (dual)
$U(1)$ into $SU(2)$, see the discussion in section~2.2. The confinement mechanism inherent
to (\ref{higgs}) is the formation of the Abrikosov-Nielsen-Olesen string
which can be considered alredy on the classical level. More generally, the Lagrangian
(\ref{higgs}) exibits the {\it Abelian dominance} in the confining region which is the dominance of
Abelian-like field configurations in the full non-Abelian theory. This dominance is common
to all the realizations of the dual-superconductor model of confinement \cite{confinement}
and is strongly supported by the lattice data \cite{review}.  What we avoid, however, is the
{\it breaking of $SU(2)$ to $U(1)$} which is inherent to the models \cite{baker} which start with
the dual gluons in the adjoint representation and then add effective isospin one Higgs fields.
Such models have well-known principal difficulties with, say,
describing interaction of the adjoint sources, see, e.g. \cite{Giacomo}.

To the contrary, the model (\ref{higgs}) can be applied to consider the static interaction of the sources
belonging to various representations of $SU(2)$. One of the basic facts here, established through
the numerical simulations on the lattice \cite{bali}, is the so called Casimir scaling. The phenomenon
of the Casimir scaling is that the static potential is described by a sum of the Coulomb-like and
linear terms:
\beq
\label{casimir}
V_j(r)~\approx~-j(j+1)\;{\alpha_s\over \pi~r}~+~j(j+1)\;\sigma\,r \,,
\eeq
where $j$ labels the representation (we consider the $SU(2)$ case) and $\sigma$ is independent of $j$.
Note that at large distances one expects qualitatively different
behavior of the potential for integer and half-integer spins $j$
because of the string breaking  in case of the integer representations. However,
at presently measured distances Eq.~(\ref{casimir}) turns to be a very good approximation to
the potential.

The potential of the type (\ref{casimir}) does arise in the classical approximation in the model (\ref{higgs})
because there are classical string solutions. However, the tension of the string is now a dynamical
quantity which can be found as a function of the parameters of the model, that is vector and Higgs masses:
\beq
\sigma ~=~ \sigma_j(\,m_H/m_V\,)\,.
\eeq
In particular, the Casimir scaling holds in the London limit, 
\beq 
{\sigma_{j_1}\over \sigma_{j_2}}~=~{j_1(j_1+1)\over j_2(j_2+1)}\,, 
\quad \mathrm{if} \quad m_H\gg m_V\,. 
\label{london} 
\eeq 
Thus, the model (\ref{higgs}) can incorporate the Casimir scaling. 

However, the description of the profile of 
the confining string is the best if $m_H\approx m_V$ \cite{ilgenfritz}.
Thus, there is mismatch with (\ref{london}).
Since the functions of $m_H/m_V$ involved in the fits
are rather smooth, it is possible to get a compromise description
which is valid in both cases, say, with 20\% accuracy.
Which is not
bad at all keeping in mind that we are using a classical
approximation. Nevertheless, the fact that the Casimir scaling                 
works at a per cent level \cite{bali} remains a kind of an unexplained
mystery in the classical approximation. Further analysis 
of this point might be needed.

\subsection{Unconventional power corrections.}

We introduced (\ref{higgs}) as an effective Lagrangian. 
Now we will describe a rather paradoxical situation that
this Lagrangian seem to provide with better phenomenology
of the power corrections to the parton approximation
than the conventional QCD approach. Namely, there are novel
$1/Q^2$ corrections inherent to Higgs models \cite{shuryak,ss}
which are absent in the standard considerations\footnote{
Literally, the model (\ref{higgs}) which introduces averaging over
all the embedding of the Abelian dual gluon into $SU(2)$ has not been
ever discussed so far. However, as far as the power corrections are
concerned there is no difference from the cases considered in 
\cite{shuryak,ss,narison}.
}. Moreover, these corrections seem to fit the data at all the 
distances measured so far, that is $r\geq 0.1 \mathrm{fm}$.
Example of this type was found in Refs. \cite{ss,narison}.
Further support came from the instanton physics \cite{shuryak}.
We have already reviewed the unconventional power corrections
in \cite{itep} and will be brief here.

In the standard approach, the power corrections are given by matrix elements
of various operators constructed on the quark and gluon fields
\cite{svz}. For our set up, the central point is that for the vacuum state
the simplest matrix element had dimension $d=4$,
\beq
\langle 0|\alpha_s(G_{\mu\nu}^a)^2|0 \rangle ~\sim ~ \Lambda_{QCD}^4\,,
\eeq
and, as a result, there are no $\Lambda_{QCD}^2/Q^2$ corrections \cite{svz}.
On the other hand, the Higgs model has a mass parameter built it.
This mass parameter can thought of as the mass of the vector particle, $m_V^2$.
Let us list a few examples where the two approaches lead to different
predictions:

(1) First, within the Higgs model the dual gluon gets mass and
the monopole potential becomes Yukawa type:
\beq
V_{m\bar{m}}~\rightarrow~{\pi\over g^2 r} e^{-m_V r}.
\eeq
The prediction has already been confirmed by the
data \cite{hoelbling}. In the conventional approach, one should have to remove at least
the term $\sim m^2r$ at short distances. The quality of the data might be not so good as to rule
this out, but the possibility looks quite bizarre.

(2) Since the dual and ``ordinary'' gluon are in fact the same particles (see discussion
in section 2) one would assume that the massiveness of the dual gluon implies
the massiveness of the gluon interacting with the color.
But this is not true \cite{ss}! There is no analyticity in this sense.
And the reason is again problems with the Dirac veto which we had
already chance to discuss in connection with the radiative corrections (see section 2.3).
Namely, as far as we discuss only the ``dual world'',
one can forget about the Dirac strings. However, if we introduce color sources $Q\bar{Q}$ into the
vacuum with $\langle\phi_M \rangle \neq 0$ we should respect the Dirac veto.
The ordinary operator product expansion, or perturbative expansion in $m_V^2/Q^2$
do not respect this veto -- like ordinary perturbation theory does not do this either
(see section 2.1). The correct treatment demonstrates that
their is a linear correction to the quark potential at short distances:
\beq
\delta V_{Q\bar{Q}}~=~\sigma_0 r\,,
\label{linear-1}
\eeq
where $\sigma_0$ is calculable function of $m_H,m_V$.
Within the standard approach there is no such term, for
explanations and further references see \cite{itep}.
The data do support the presence of the linear term.
Amusingly enough, the data refer exclusively to the
the nonperturbative potential, and there is no need for
painful separation of (small) power corrections against the
perturbative ``background''.
 
(3) The linear term (\ref{linear-1}) can be rephrased as the statement
that the gluon has a tachyonic mass. Indeed, the ordinary mass would give a negative $\sigma_0$,
as seen from the expansion of the Yukawa potential at short distances.
The introduction of a tachyonic gluon mass in the framework of the QCD sum
rules allows to resolve in an absolutely natural way long standing problems with the
phenomenology based on the QCD sum rules \cite{novikov}.
 
 (4) The last but not the least point in our discussion concerns the instanton density \cite{shuryak}.
The conventional approach predicts that the deviations from the 't~Hooft instanton
density due to non-trivial background vacuum fields is
of the fourth order in the instanton size $\rho$: 
\beq
dn(\rho)~=~dn_{pert.}(\rho)\,(1~+~ 
\frac{\pi^4 \rho^4}{2 g^4} \,\langle 0| g^2 (G_{\mu\nu}^a)^2|0 \rangle ~+~ ...)
\eeq
The data, on the other hand are beautifully fitted by a quadratic correction,
inherent to (\ref{higgs}). Note however, that the coefficient
in front of the quadratic term has been fitted rather than calculated
from (\ref{higgs}) so far.

\subsection {Conclusions  \# 4}

We have proposed in this section a phenomenological Lagrangian (\ref{higgs}) which unifies 
the Higgs mechanism for the Abelian dual gluon with full $SU(2)$ symmetry
of the ordinary gluodynamics. The full study of the consequences 
from this formulation is still awaiting its time to come.

However, it seems promising that the color SU(2) is not broken at any
step despite the Higgs mechanism.
This allows to broaden applications of the effective Lagrangian and incorporate, to certain
accuracy, the Casimir scaling. 

Also, emergence of the mass of the dual gluon in the effective Lagrangian approach provides
with a natural framework to introduce the novel $1/Q^2$ corrections. Phenomenologically,
these corrections bring crucial improvements to the existing phenomenology. Moreover,
generically the corrections are of the same type as those associated with ultravioloet
renormalons (see, e.g., \cite{akhoury-zakharov}). However, within the effective Lagrangian
approach these corrections should disappear in the limit of infinite $Q^2$
which is not true for the ultraviolet renormalons and has not been
supported by any data so far.

\section{Conclusions.}

In this review we considered various effects related to the monopoles in unbroken non-Abelian gauge theories.
In conclusion, let us reiterate the main poins (see also conclusions to the Sections 1--4):

{\it (i)} Fundamental monopoles with the magnetic charge $|Q_m|=1$ are introduced
as external objects via the 't~Hooft loop. The corresponding intermonopole potential $V_{m\bar{m}}(r)$
can be evaluated at short distances from first principles.  In the Lagrangian approach similar to that
of Zwanziger, the dual gluon interacting with point-like external monopoles appears as an {\it Abelian gauge field} 
(see Section 2 and \cite{main,main2} for details).

{\it (ii)} Monopoles with the magnetic charge $|Q_m|=2$ are pure quantum objects which can be studied
so far only numerically. We discussed briefly the newly intoduced \cite{main1}
geometrical monopoles which appear to be physical objects.

{\it (iii)}  Effective Lagrangian which assumes condensation of the monopoles incorporates the Abelian dominance
at distances, where the effects of confinement are crucial, without breaking $SU(2)$ to $U(1)$. 
In the London limit, it reproduces the the Casimir scaling phenomenon.
There are further phenomenological consequences, in particular,  the evaluation of the
potential $V_{m\bar{m}}(r)$ at larger distances (see Section 4 and \cite{main,main2} for details).

\section*{Acknowledgments} 
\noindent 
The review is based to a large extent on talks presented by the authors on various conferences this year.
The authors are gratefull to the organizers for the invitations.
We are thankful to V.A.~Shevchenko, Yu.A.~Simonov, L.~Stodolsky, T.~Suzuki, V.A.~Rubakov for
valuable discussions. 
M.N.Ch. and M.I.P. acknowledge the kind hospitality of the staff of the
Max-Planck Institut f\"ur Physik (M\"unchen), where the work was initiated.  Work of M.N.C.,
F.V.G. and M.I.P. was partially supported by grants RFBR 99-01230a and INTAS 96-370.
M.N.Ch. and M.I.P. are supported by Monbushu grant and CRDF award RP1-2103.

\end{document}